\documentclass[10pt,twoside,english,prl,aps,twocolumn,twoside,superscriptaddress]{revtex4-1}

\usepackage[T1]{fontenc}
\usepackage[latin9]{inputenc}
\usepackage{color}
\usepackage{babel}
\usepackage{mathtools}
\usepackage{amsmath}
\usepackage{amsthm}
\usepackage{amssymb}
\usepackage{cancel}
\usepackage{mathdots}
\usepackage{stmaryrd}
\usepackage{stackrel}
\PassOptionsToPackage{version=3}{mhchem}
\usepackage{mhchem}
\usepackage{esint}
\usepackage[unicode=true,pdfusetitle,
 bookmarks=true,bookmarksnumbered=false,bookmarksopen=false,
 breaklinks=true,pdfborder={0 0 0},pdfborderstyle={},backref=false,colorlinks=true]
 {hyperref}
\hypersetup{
 urlcolor=red, citecolor=blue}

\makeatletter
\theoremstyle{plain}
\newtheorem{thm}{\protect\theoremname}
\theoremstyle{plain}
\newtheorem{cor}{\protect\corollaryname}
\theoremstyle{plain}
\newtheorem*{assumption*}{\protect\assumptionname}

\usepackage{babel}
\SetSymbolFont{stmry}{bold}{U}{stmry}{b}{n}

\makeatother

\providecommand{\assumptionname}{Assumption}
\providecommand{\corollaryname}{Corollary}
\providecommand{\theoremname}{Theorem}

\begin{document}

\title{Impossibility of creating a superposition of unknown quantum states }

\author{Somshubhro Bandyopadhyay }
\email{som@jcbose.ac.in; som.s.bandyopadhyay@gmail.com}

\affiliation{Department of Physics, Bose Institute, Unified Academic Campus, EN
80, Sector V, Bidhannagar, Kolkata 700091, India}
\begin{abstract}
The superposition principle is fundamental to quantum theory. Yet
a recent no-go theorem has proved that quantum theory forbids superposition
of unknown quantum states, even with nonzero probability. The implications
of this result, however, remain poorly understood so far. In this
paper we show that the existence of a protocol that superposes two
unknown pure states with nonzero probability (allowed to vary over
input states) leads to the violation of other no-go theorems. In particular,
such a protocol can be used to perform certain state discrimination
and cloning tasks that are forbidden not only in quantum theory but
in no-signaling theories as well. 
\end{abstract}
\maketitle
In quantum theory, a state of a physical system is a vector $\left|\psi\right\rangle $
of unit norm in a Hilbert space $\mathcal{H}$. Furthermore, $\left|\psi\right\rangle $
and $e^{i\vartheta}\left|\psi\right\rangle $ describe the same physical
state of the system, where $\left|e^{i\vartheta}\right|=1$; thus,
a global phase is inconsequential. The ``superposition principle''
states that for any two vectors $\left|\psi\right\rangle ,\left|\phi\right\rangle \in\mathcal{H}$
and nonzero complex numbers $\gamma,\delta$ satisfying $\left|\gamma\right|^{2}+\left|\delta\right|^{2}=1$,
the linear superposition $\gamma\left|\psi\right\rangle +\delta\left|\phi\right\rangle \in\mathcal{H}$
is also a state of the system under consideration. However, unlike
a global phase, the relative phase in a superposition is physically
significant, i.e. $\gamma\left|\psi\right\rangle +\delta\left|\phi\right\rangle $
and $\gamma\left|\psi\right\rangle +\delta e^{i\vartheta}\left|\phi\right\rangle $
represent two different states of the same physical system. The superposition
principle is fundamental to quantum theory. In fact, almost all nonclassical
properties exhibited by quantum systems, e.g., nonorthogonality of
quantum states \citep{quantum}, quantum interference \citep{quantum,Zeilinger-1999},
quantum entanglement \citep{quantum,entanglement}, and quantum coherence
\citep{coherence} are consequences of quantum superpositions. 

Recently a basic question, closely related to quantum superposition,
was considered \citep{no-superposition}: Does there exist a quantum
operation that would superpose two unknown pure quantum states with
some complex weights? The question is of particular interest because
quantum theory is known to forbid physical realizations of certain
operations, even plausible ones \citep{no-cloning-WZ,no-cloning-2,no-broadcasting,no-disentangling,no-deleting,Pati-2002,no-local-broadcasting,no-coherence,Araujo+-2014,Thompson+2018},
and therefore, it is important to understand whether similar restrictions
are also in place on something as basic as the creation of quantum
superpositions. Besides, exploring such questions often reveals new
ways of manipulating quantum systems that have found useful applications
in quantum information and computation. 

Before we proceed, we note here that a special case of the above question
was initially posed in Ref.$\,$\citep{no-adder} where the authors
asked about the existence of a quantum adder, a unitary operator that
would add two unknown pure quantum states, and proved that such a
unitary operator cannot exist. The proof followed from the observation
that an unobservable global phase associated with the input state
can distribute itself in infinitely many ways in a superposition,
thereby leading to infinitely many superpositions with observable
relative phases, which is unphysical.

Let us now consider the general formulation of the question on the
existence of quantum superposers \citep{no-superposition}: For given
nonzero complex numbers $\alpha,\beta$ satisfying $\left|\alpha\right|^{2}+\left|\beta\right|^{2}=1$
and any given pair of vectors $\left|\psi\right\rangle ,\left|\phi\right\rangle \in\mathcal{H}$,
does there exist a quantum protocol that prepares the superposed state
$\left|\Psi\right\rangle \propto\alpha\left|\psi\right\rangle +\beta\left|\phi\right\rangle $?
The ambiguity of the relative phase, which ruled out the existence
of a quantum adder, however, is also present in this general formulation.
Let $\rho_{\chi}=\left|\chi\right\rangle \left\langle \chi\right|$
denote the density matrix for any normalized vector $\left|\chi\right\rangle $.
Then it is easy to see that $\rho_{\Psi}$ cannot be a well-defined
function of $\rho_{\psi}$ and $\rho_{\phi}$ for the simple reason
that the density matrix $\rho_{\chi}$ corresponds not only to $\left|\chi\right\rangle $
but also to any other normalized vector $\left|\chi^{\prime}\right\rangle =e^{i\theta}\left|\chi\right\rangle $.
The authors \citep{no-superposition} therefore relaxed the definition
of superposing such that there's no phase ambiguity. Specifically,
for any pair of vectors $\left|\psi\right\rangle ,\left|\phi\right\rangle \in\mathcal{H}$
they allowed for complex superpositions of any two vectors with density
matrices $\rho_{\psi}$ and $\rho_{\phi}$. With this, the question
becomes well defined. Then a superposition protocol, if one such exists,
could be realized by application of a quantum channel on the input
systems and then tracing out one of them. The authors also allowed
post-selection, which entails the possibility of obtaining the desired
output with some nonzero probability. In other words, the most general
class of quantum operations, described in terms of trace-nonincreasing
completely positive (CP) maps, was considered in Ref.$\,$\citep{no-superposition}.
The answer, however, turned out to be no. 
\begin{thm}
\citep{no-superposition} Let $\alpha,\beta$ be any two nonzero complex
numbers satisfying $\left|\alpha\right|^{2}+\left|\beta\right|^{2}=1$.
Let $\mathcal{H}$ be a Hilbert space, where $\dim\mathcal{H}\geq2$.
Then there does not exist a nonzero CP map $\Lambda_{\alpha,\beta}:\mathcal{H}^{\otimes2}\rightarrow\mathcal{H}$
such that for all pure states $\rho_{\psi},\rho_{\phi}\in\mathcal{H}$,
$\Lambda_{\alpha,\beta}\left(\rho_{\psi}\otimes\rho_{\phi}\right)\propto\left|\Psi\right\rangle \left\langle \Psi\right|,$
where $\left|\Psi\right\rangle =\alpha\left|\psi\right\rangle +\beta\left|\phi\right\rangle $
and the states appearing in the superposition may in general depend
on both $\rho_{\psi}$ and $\rho_{\phi}$. 
\end{thm}
Noting that the superposition in general may depend on both $\rho_{\psi}$
and $\rho_{\phi}$ and a global phase is not of any consequence, one
has the following corollary. 
\begin{cor}
Let $\alpha,\beta$ be any two nonzero complex numbers satisfying
$\left|\alpha\right|^{2}+\left|\beta\right|^{2}=1$. Let $\mathcal{H}$
be a Hilbert space, where $\dim\mathcal{H}\geq2$. Then there does
not exist a nonzero CP map $\Lambda_{\alpha,\beta}:\mathcal{H}^{\otimes2}\rightarrow\mathcal{H}$
such that for all pure states $\left|\psi\right\rangle ,\left|\phi\right\rangle \in\mathcal{H}$,
$\Lambda_{\alpha,\beta}\left(\rho_{\psi}\otimes\rho_{\phi}\right)\propto\left|\Psi\right\rangle \left\langle \Psi\right|,$
where $\left|\Psi\right\rangle =\alpha\left|\psi\right\rangle +\beta e^{i\theta}\left|\phi\right\rangle $
for some phase $\theta\in\left[0,2\pi\right)$ that may in general
depend on the input states. 
\end{cor}
The above result, known as the no-superposition theorem, forbids the
existence of a universal probabilistic quantum superposer: a quantum
operation that would superpose two unknown pure quantum states with
nonzero probability. The result forms yet another no-go theorem in
quantum theory. 

Now the no-go theorems \citep{no-cloning-WZ,no-cloning-2,no-broadcasting,no-disentangling,no-deleting,Pati-2002,no-local-broadcasting,no-coherence,Araujo+-2014,Thompson+2018}
in quantum theory are of particular significance because they tell
us which operations are physically allowed and which are not. For
example, the no-cloning theorem \citep{no-cloning-WZ} states that
it is impossible to make exact copies of an unknown quantum state.
But at a more fundamental level, the no-go theorems can have deep
implications. For example, in a world without the no-cloning theorem,
it is possible to send signals faster than light \citep{clong-signaling,Hardy-Song}
and reliably distinguish nonorthogonal states, both of which would
lead to a complete breakdown of our existing physical theories. So
while the no-go theorems are fairly easy to understand, their implications
can be far reaching, but often not immediate. 

The implications of the no-superposition theorem, however, remain
poorly understood so far. Neither do we know of any relation with
any other existing no-go result, nor do we know of the consequences,
if any, should it be violated. Although follow-up papers have come
up with interesting results \citep{Dogra+-2018,Doosti+-2017,Sami+-2016}
and variants \citep{Luo+-2017}, none could account for the most basic
questions: Why is it not possible to superpose unknown quantum states,
even with a nonzero probability? And what would be the consequences
if we could? 

In this paper, we will show that the existence of universal probabilistic
quantum superposers implies the existence of protocols that can perform
certain quantum state discrimination and cloning tasks forbidden not
only in quantum theory, but also in no-signaling theories. So indeed,
there will be unphysical consequences should such quantum superposers
exist. 

We begin by assuming that universal probabilistic quantum superposers
exist. That is, 
\begin{assumption*}
For every pair of nonzero complex numbers $\alpha,\beta$ satisfying
$\left|\alpha\right|^{2}+\left|\beta\right|^{2}=1$, there exists
a universal probabilistic quantum superposer $\mathcal{Q}_{\alpha,\beta}$
that for any two pure quantum states $\left|\psi\right\rangle ,\left|\phi\right\rangle $
prepares, with probability $p_{\psi,\phi}^{\alpha,\beta}>0$, a superposition
state $\left|\Psi\right\rangle \propto\alpha\left|\psi\right\rangle +\beta e^{i\theta}\left|\phi\right\rangle $
for some phase $\theta\in\left[0,2\pi\right)$, where $\theta$ may
in general depend on the input states. 
\end{assumption*}
Thus $\mathcal{Q}_{\alpha,\beta}$ is a two-input, single-output,
quantum black box that takes a pair of pure quantum states as input
and generates their linear superposition as output with some nonzero
probability which is allowed to vary over input states (for the sake
of full generality).

The basic idea is to show that the existence of $\mathcal{Q}_{\alpha,\beta}$
implies violation of the following theorems: 

Let $S_{\psi}=\left\{ \left|\psi_{1}\right\rangle ,\left|\psi_{2}\right\rangle ,\dots,\left|\psi_{n}\right\rangle \right\} $
be a set of pure states such that $0\leq\left|\left\langle \psi_{i}\vert\psi_{j}\right\rangle \right|<1$
for $i\neq j$. Then, 
\begin{itemize}
\item the states can be unambiguously distinguished (i.e., every state in
the set can be correctly identified with a nonzero probability) if
and only if they are linearly independent \citep{Chefles-unambiguous}. 
\item the states can be probabilistically cloned if and only if they are
linearly independent \citep{prob-clone}. 
\end{itemize}
Note that the above two statements are equivalent in the following
sense: For any given set of states, if unambiguous discrimination
is possible, then probabilistic cloning is possible as well and vice
versa. Further note that the constraint on probabilistic cloning of
states follows from the condition of no faster-than-light signaling
\citep{Hardy-Song}. So quantum theory, and independently, the no-signaling
condition, forbid unambiguous discrimination and probabilistic cloning
of linearly dependent pure states. This in turn implies the following: 
\begin{itemize}
\item Let $S_{\psi}=\left\{ \left|\psi_{1}\right\rangle ,\left|\psi_{2}\right\rangle ,\dots,\left|\psi_{n}\right\rangle \right\} $
be a set of linearly dependent pure states. Then, there does not exist
a quantum protocol that achieves the state transformation $\left|\psi_{i}\right\rangle \rightarrow\left|\Psi_{i}\right\rangle $
for every $i$ with a nonzero probability such that the states $\left|\Psi_{1}\right\rangle ,\left|\Psi_{2}\right\rangle ,\dots,\left|\Psi_{n}\right\rangle $
are linearly independent. 
\end{itemize}
The proof is simple. Suppose that a quantum system is prepared in
a state chosen from the known set $S_{\psi}$ but we do not know which
state. Now as the states $\left|\psi_{i}\right\rangle $ are linearly
dependent, quantum theory will not allow us to correctly identify
the state of the system, or make copies of it, even with nonzero probability.
But it is easy to see that both the tasks become possible if there
exists a protocol that achieves the transformation $\left|\psi_{i}\right\rangle \rightarrow\left|\Psi_{i}\right\rangle $
with nonzero probability for every $i$, where the states $\left|\Psi_{1}\right\rangle ,\left|\Psi_{2}\right\rangle ,\dots,\left|\Psi_{n}\right\rangle $
are linearly independent. Therefore such a protocol can not exist. 

Let us now consider a $\mathcal{Q}_{\alpha,\beta}$-based state transformation
protocol that works as follows. We feed our quantum machine $\mathcal{Q}_{\alpha,\beta}$
with two input states: The first is chosen from a known set $S_{\psi}=\left\{ \left|\psi_{1}\right\rangle ,\left|\psi_{2}\right\rangle ,\dots,\left|\psi_{n}\right\rangle \right\} $
of pure states and the second is some pure state $\left|\phi\right\rangle $.
In this way it is possible to prepare an ensemble $S_{\Psi}=\left\{ \left|\Psi_{1}\right\rangle ,\left|\Psi_{2}\right\rangle ,\dots,\left|\Psi_{n}\right\rangle \right\} $
of output states, where each state $\left|\Psi_{j}\right\rangle \propto\alpha\left|\psi_{j}\right\rangle +\beta e^{i\theta_{j}}\left|\phi\right\rangle $
is generated with some nonzero probability $p_{\psi_{i},\phi}^{\alpha,\beta}$.
Thus we have a protocol that transforms $\left|\psi_{j}\right\rangle \rightarrow\left|\Psi_{j}\right\rangle $
with a nonzero probability for every $j$. 

For a given $\mathcal{Q}_{\alpha,\beta}$ and $S_{\psi}$, observe
that the output states will be different for different choices of
$\left|\phi\right\rangle $. To make this explicit, denote the set
of output states by $S_{\Psi}\left(\phi\right)$. So if the states
$\left|\psi_{1}\right\rangle ,\left|\psi_{2}\right\rangle ,\dots,\left|\psi_{n}\right\rangle $\emph{
}are linearly dependent,\emph{ }then for every $\left|\phi\right\rangle $
the states in $S_{\Psi}\left(\phi\right)$ must also be linearly dependent
because otherwise, the protocol would be unphysical. 

We now give a simple proof that the protocol, in fact, is unphysical
for all $\mathcal{Q}_{\alpha,\beta}$. Consider a set of three linearly
dependent pure states that belong to a $d$-dimensional Hilbert space,
where $d\geq3$. The states are given by
\begin{eqnarray}
\left|\psi_{1}\right\rangle  & = & \left|\psi\right\rangle ,\nonumber \\
\left|\psi_{2}\right\rangle  & = & \left|\psi^{\perp}\right\rangle ,\label{psi}\\
\left|\psi_{3}\right\rangle  & = & a\left|\psi\right\rangle +b\left|\psi^{\perp}\right\rangle ,\nonumber 
\end{eqnarray}
where $a,b\neq0$, $a,b\in\mathbb{R}$, $a^{2}+b^{2}=1$. By construction,
the states are linearly dependent as they live in the two-dimensional
subspace spanned by $\left\{ \left|\psi\right\rangle ,\left|\psi^{\perp}\right\rangle \right\} $. 

Following the protocol, we feed $\mathcal{Q}_{\alpha,\beta}$ with
two input states, where the first input is chosen from $\left\{ \left|\psi_{1}\right\rangle ,\left|\psi_{2}\right\rangle ,\left|\psi_{3}\right\rangle \right\} $
as given above, and the second input state is taken to be a pure state
$\left|\phi\right\rangle $ which is orthogonal to both $\left|\psi\right\rangle $
and $\left|\psi^{\perp}\right\rangle $. Then the possible output
states, each of which is generated with some nonzero probability,
are given by 
\begin{eqnarray}
\left|\Psi_{1}\right\rangle  & = & \alpha\left|\psi_{1}\right\rangle +\beta e^{i\theta_{1}}\left|\phi\right\rangle ,\nonumber \\
\left|\Psi_{2}\right\rangle  & = & \alpha\left|\psi_{2}\right\rangle +\beta e^{i\theta_{2}}\left|\phi\right\rangle ,\label{Psi}\\
\left|\Psi_{3}\right\rangle  & = & \alpha\left|\psi_{3}\right\rangle +\beta e^{i\theta_{3}}\left|\phi\right\rangle .\nonumber 
\end{eqnarray}
We will show that the states $\left|\Psi_{j}\right\rangle $ for $j=1,2,3$
are linearly independent; that is, the equation
\begin{eqnarray}
x_{1}\left|\Psi_{1}\right\rangle +x_{2}\left|\Psi_{2}\right\rangle +x_{3}\left|\Psi_{3}\right\rangle  & = & 0\label{LI-condition}
\end{eqnarray}
holds if and only if $x_{1}=x_{2}=x_{3}=0$. The \emph{if} part is
trivial. So let us now consider the \emph{only if} part. 

First, we write Eq.$\,$(\ref{LI-condition}) as 
\begin{eqnarray}
\alpha\left(x_{1}+ax_{3}\right)\left|\psi\right\rangle +\alpha\left(x_{2}+bx_{3}\right)\left|\psi^{\perp}\right\rangle +\beta\sum_{j=1}^{3}e^{i\theta_{j}}x_{j}\left|\phi\right\rangle  & = & 0\nonumber \\
\label{LI-condition-1}
\end{eqnarray}
As the states $\left|\psi\right\rangle ,\left|\psi^{\perp}\right\rangle ,\left|\phi\right\rangle $
are mutually orthogonal they are linearly independent. Thus the coefficients
appearing in the above superposition must vanish, i.e., 
\begin{eqnarray}
\alpha\left(x_{1}+ax_{3}\right) & = & 0,\label{alpha,x,z}\\
\alpha\left(x_{2}+bx_{3}\right) & = & 0,\label{alpha,y,z}\\
\beta\left(e^{i\theta_{1}}x_{1}+e^{i\theta_{2}}x_{2}+e^{i\theta_{3}}x_{3}\right) & = & 0.\label{beta,x,y,z}
\end{eqnarray}
Since $\alpha,\beta\neq0$, we have 
\begin{eqnarray}
x_{1}+ax_{3} & = & 0,\label{x,z}\\
x_{2}+bx_{3} & = & 0,\label{y,z}\\
e^{i\theta_{1}}x_{1}+e^{i\theta_{2}}x_{2}+e^{i\theta_{3}}x_{3} & = & 0.\label{x,y,z,theta}
\end{eqnarray}
Let us now find the conditions under which the above three equations
are satisfied simultaneously. As $a,b\neq0$, we see that the above
three equations are simultaneously satisfied when $x_{i}=0$ for all
$i=1,2,3$. We will now show that this is the only solution. To establish
this, assume that $x_{i}\neq0$ for all $i=1,2,3$. Then from (\ref{x,z})
and (\ref{y,z}) we get $x_{1}=-ax_{3}$, and $x_{2}=-bx_{3}$. Substituting
these in (\ref{x,y,z,theta}) and noting that $x_{3}\neq0$ we obtain
\begin{eqnarray}
e^{i\theta_{1}}a+e^{i\theta_{2}}b-e^{i\theta_{3}} & = & 0,\label{a,b,thetas}
\end{eqnarray}
or, equivalently, 
\begin{eqnarray}
a+e^{i\theta_{21}}b & = & e^{i\theta_{31}},\label{a,b,theta21,theta31}
\end{eqnarray}
where $\theta_{21}=\theta_{2}-\theta_{1}$ and $\theta_{31}=\theta_{3}-\theta_{1}$.
Equation (\ref{a,b,theta21,theta31}) implies that 
\begin{eqnarray}
\left|a+b^{\prime}\right| & = & 1,\label{|a+bprime|=00003D1}
\end{eqnarray}
where $b^{\prime}=e^{i\theta_{21}}b$. Now, recall that $a,b\neq0$,
$a,b\in\mathbb{R}$, and $a^{2}+b^{2}=1$. Then 
\begin{eqnarray}
a^{2}+\left|b^{\prime}\right|^{2} & = & 1.\label{a^2+|bprime|^2=00003D1}
\end{eqnarray}
A simple calculation shows that Eqns.$\,$(\ref{|a+bprime|=00003D1})
and (\ref{a^2+|bprime|^2=00003D1}) are satisfied only when $\theta_{21}=\pi/2,3\pi/2$
(since, $a,b\neq0$). Then from (\ref{a,b,theta21,theta31}) it follows
that $a=\cos\theta_{31}$ and \textbf{$b=\pm\sin\theta_{31}$}. But
the phases that $\mathcal{Q}_{\alpha,\beta}$ associate with the superposition
cannot have any dependence on $a$ and $b$! This is because $a$
and $b$ are basis-dependent coefficients and $\left|\psi_{3}\right\rangle $
has infinitely many such representations. In other words, while $\theta_{3}$
may depend on $\left|\psi_{3}\right\rangle $ and $\left|\phi\right\rangle $,
it cannot depend on the basis representation of $\left|\psi_{3}\right\rangle $.
So the solutions are not feasible. 

Therefore, Eqns.$\,$ (\ref{x,z}), (\ref{y,z}), and (\ref{x,y,z,theta})
can only be simultaneously satisfied when $x_{i}=0$, $i=1,2,3.$
Consequently, the only solution to (\ref{LI-condition}) is $x_{1}=x_{2}=x_{3}=0$.
Thus the states $\left|\Psi_{1}\right\rangle ,\left|\Psi_{2}\right\rangle ,\left|\Psi_{3}\right\rangle $
are linearly independent. Note that the analysis holds for all $\mathcal{Q}_{\alpha,\beta}$.
This completes the proof. 

Thus we have shown that the existence of $\mathcal{Q}_{\alpha,\beta}$
implies the existence of a protocol that can transform linearly dependent
pure states into linearly independent pure states. As explained earlier,
such protocols can be used to perform the tasks of unambiguous discrimination
and probabilistic cloning of linearly dependent pure states, both
of which are forbidden in quantum theory and also violate the no-signaling
condition. So the unconditional superposition of unknown quantum states,
even probabilistically, gives rise to unphysical consequences. 

Now, interestingly, there exists a probabilistic quantum protocol
to superpose two unknown pure states \citep{no-superposition}. Not
surprisingly though, the protocol comes with strings attached --
in particular, each input state must have a fixed overlap with a known
reference state and the superposition coefficients are functions of
the overlaps as well. It is now clear why these conditions are necessary,
for without these constraints probabilistic superposition would not
be possible. 

Our result also sheds light on a recent theorem \citep{Luo+-2017}
which states that it is possible to superpose two unknown pure states
chosen from a known set if and only if the states are linearly independent.
The \emph{if} part here is easy to understand: If the states are linearly
independent, then in the first step, one correctly identifies the
input states by performing suitable unambiguous state discrimination
measurements and in the next step creates the desired superposition.
The\emph{ only if} part, on the other hand, can be understood by noting
that $Q_{\alpha,\beta}$ can transform linearly dependent pure states
into linearly independent ones, which is unphysical. This implies
the states in the given set must all be linearly independent. 

\emph{Conclusions. }The no-go theorems in quantum theory help us to
understand the class of allowed physical operations. But it is also
equally important to understand the consequences should no-go theorems
be violated and the answers must come from physics. Here we showed
that the existence of a protocol that superposes unknown pure states,
even with nonzero probability, leads to unambiguous discrimination
and probabilistic cloning of linearly dependent pure states -- tasks
that are forbidden in quantum theory and also in no-signaling theories. 

One question, in the context of the present result, however, remains
open: What kind of unphysical consequences would arise if a universal
probabilistic quantum superpower, assuming it exists, is allowed to
admit only qubit states? We could not find a satisfactory answer.
Nevertheless, we are hopeful that a satisfactory answer will eventually
be found, perhaps considering a different physical scenario.
\begin{acknowledgments}
The author is grateful to Guruprasad Kar and Tomasz Paterek for many
helpful discussions. 
\end{acknowledgments}

\end{document}